\documentclass[12pt,twoside,dvips]{article}
\usepackage{amsmath}
\usepackage[spanish,english]{babel}
\usepackage{graphicx}

\begin{document}

\title{Family Dependence in $SU(3)_{c}\otimes SU(3)_{L}\otimes U(1)_{X}$
models}
\author{Fredy Ochoa$\thanks{%
e-mail: faochoap@unal.edu.co}$ and R. Mart\'{\i}nez$\thanks{%
e-mail: remartinezm@unal.edu.co}$ \and Departamento de F\'{\i}sica,
Universidad Nacional, \\
Bogot\'{a}-Colombia}
\maketitle

\begin{abstract}
Using experimental results at the $Z$-pole and atomic parity violation, we
perform a $\chi ^{2}$ fit at 95\% CL to obtain family-dependent bounds to $%
Z_{2}$ mass and $Z_{\mu }-Z_{\mu }^{\prime }$ mixing angle $\theta $ in the
framework of $SU(3)_{c}\otimes SU(3)_{L}\otimes U(1)_{X}$ models. The
allowed regions depend on the assignment of the quark families in mass
eigenstates into the three different families in weak eigenstates that
cancel anomalies.
\end{abstract}

\vspace{-0.3cm}

\vspace{-5mm}

\section{Introduction}

A very common alternative to solve some of the problems of the standard
model (SM) consists on enlarging the gauge symmetry. For instance, the $%
SU(5) $\ grand unification model of Georgi and Glashow \cite{seven} can
unify the interactions and predicts the electric charge quantization; while
the group $E_{6}$ can also unifies the interactions and might explain the
masses of the neutrinos \cite{nine}. Nevertheless, such models cannot
explain the origin of the fermion families. Some models with larger
symmetries address this problem \cite{familia, Frampton2}. A very
interesting alternative to explain the origin of generations comes from the
cancellation of chiral anomalies \cite{anomalias}. In particular, models
with gauge symmetry $SU(3)_{c}\otimes SU(3)_{L}\otimes U(1)_{X},$ also
called 331 models, arise as a possible solution to this puzzle, where the
three families are required in order to cancel chiral anomalies completely.
An additional motivation to study these kind of models comes from the fact
that they can also predict the charge quantization for a three family model
even when neutrino masses are added \cite{Pires}.

Although cancellation of anomalies leads to some required conditions \cite%
{fourteen}, such criterion alone still permits an infinite number of 331
models. In these models, the electric charge is defined in general as a
linear combination of the diagonal generators of the group 
\begin{equation}
Q=T_{3}+\beta T_{8}+XI.  \label{charge}
\end{equation}%
As it has been studied in the literature \cite{fourteen, ten}, the value of
the $\beta $ parameter determines the fermion assignment, and more
specifically, the electric charges of the exotic spectrum. Hence, it is
customary to use this quantum number to classify the different 331 models.
If we want to avoid exotic charges we are led to only two different models
i.e. $\beta =\pm 1/\sqrt{3}$ \cite{fourteen, twelve}.

It has been recently obtained in ref \cite{331us} constraints to 331 models
by examining the scalar sector. In summary, these constraints are obtained
by requiring gauge invariance in the Yukawa sector and finding the possible
vacuum alignment structures that respect the symmetry breaking pattern and
provides the fermions and gauge bosons of the SM with the appropriate
masses. By applying gauge invariance to the Yukawa lagrangian it is found
that the Higgs bosons should lie in either a triplet, antitriplet, singlet
or sextet representation of $SU\left( 3\right) _{L}$. On the other hand,
cancellation of chiral anomalies demands that the number of fermionic
triplets and antitriplets must be equal \cite{doff}. Moreover, assuming the
symmetry breaking pattern $SU(3)_{L}\otimes U(1)_{X}\rightarrow $ $%
SU(2)_{L}\otimes U(1)_{Y}\rightarrow U(1)_{Q},$ we see that one triplet is
necessary for the first symmetry breaking and two triplets for the second in
order to give mass to the up and down quark sector. In some cases it is
necessary to introduce a scalar sextet to give masses to the neutrinos \cite%
{ten}.

The group structure of these models leads, along with the SM neutral boson $%
Z,$ to the prediction of an additional current associated with a new neutral
boson $Z^{\prime }.$ Unlike $Z$-boson whose couplings are family independent
and the weak interactions at low energy are of universal character, the
couplings of $Z^{\prime }$ are different for the three families due to the $%
U(1)_{X}$ values to each of them. Through the $Z-Z^{\prime }$ mixing it is
possible to study the low energy deviations of the $Z$ couplings to the SM
families \cite{ten, Carena, Frampton}. In the quark sector each 331-family
in the weak basis can be assigned in three different ways into mass
eigenstates. In this way in a phenomenological analysis, the allowed region
associated with the $Z-Z^{\prime }$ mixing angle and the physical mass $%
M_{Z_{2}}$ of the extra neutral boson will depend on the family assignment
to the mass states.

In this work we report a phenomenological study through a $\chi ^{2}$ fit at
the $Z$-pole to find the allowed region for the mixing angle between the
neutral gauge bosons $Z-Z^{\prime }$ and the mass of the $Z_{2}$ boson at
95\% CL for three different assignments of the quark families \cite%
{Mohapatra}. We take into account the two main versions of the 331 models
from the literature \cite{ten, twelve}, which correspond to $\beta =-\sqrt{3}
$ and $-\frac{1}{\sqrt{3}}$ respectively.

This paper is organized as follows. Section \ref{sec:331-spectrum} is
devoted to summarize the Fermion, Scalar and Vector boson representations.
In section \ref{sec:neutral-currents} we describe the neutral currents and
the vector and axial vector couplings of the model. In section \ref%
{sec:z-pole} we perform the $\chi ^2$ analysis at the Z-pole including
atomic parity violation at 95\% CL. Finally, section \ref{conclusions}
contains our conclusions.

\section{The 331 spectrum\label{sec:331-spectrum}}

The fermionic spectrum under $SU(3)_{c}\otimes $ $SU(3)_{L}\otimes U(1)_{X}$
is shown in table \ref{tab:espectro} for three families with $\beta =-\frac{1%
}{\sqrt{3}}$ and $-\sqrt{3},$ where the first case contains the Long model 
\cite{twelve} and the second contains the bilepton model proposed by Pisano,
Pleitez and Frampton \cite{ten}. We recognize three different possibilities
to assign the physical quarks in each family representation as it is shown
in table \ref{tab:combination}. At low energy, the three models from table %
\ref{tab:combination} are equivalent and there are not any phenomenological
feature that allow us to detect differences between them. In fact, they must
reduce to the SM which is an universal family model in $SU(2)_{L}.$ However,
through the couplings of the three families to the additional neutral
current ($Z^{\prime }$) and the introduction of a mixing angle between $Z$
and $Z^{\prime }$ it is possible to recognize differences among the three
models at the electroweak scale. It is noted that although we write the
spectrum in the weak basis in table \ref{tab:espectro}, we can consider
three realizations in the mass basis in table \ref{tab:combination}.

\begin{table}[tbp]
\begin{equation*}
\begin{tabular}{||c||}
\hline\hline
Representation \\ \hline\hline
$%
\begin{tabular}{c}
$q_{mL}=\left( 
\begin{array}{c}
d_{m} \\ 
-u_{m} \\ 
J_{m}^{-1/3\text{ }(-4/3)}%
\end{array}%
\right) _{L}:\left( \mathbf{3,3}^{\ast },0\text{ }\left( -1/3\right) \right)
_{-1/\sqrt{3}\text{ }(-\sqrt{3})}$ \\ 
\\ 
$d_{mR},$ $u_{mR},$ $J_{mR}^{-1/3\text{ }(-4/3)}:\mathbf{1}$%
\end{tabular}%
\ \ $ \\ \hline\hline
\begin{tabular}{c}
$q_{3L}=\left( 
\begin{array}{c}
u_{3} \\ 
d_{3} \\ 
J_{3}^{2/3\text{ }(5/3)}%
\end{array}%
\right) _{L}:\left( \mathbf{3,3},1/3\text{ (}2/3\text{)}\right) _{-1/\sqrt{3}%
\text{ (}-\sqrt{3})}$ \\ 
\\ 
$u_{3R},$ $d_{3R},$ $J_{3R}^{2/3\text{ }(5/3)}:\mathbf{1}$%
\end{tabular}
\\ \hline\hline
\begin{tabular}{c}
$\ell _{jL}=\left( 
\begin{array}{c}
\nu _{j} \\ 
e_{j} \\ 
E_{j}^{0\text{ (}1)}%
\end{array}%
\right) _{L}:\left( \mathbf{3,3},-1/3\text{ }\left( 0\right) \right) _{-1/%
\sqrt{3}\text{ (}-\sqrt{3})}$ \\ 
\\ 
$e_{jR},$ $E_{jR}^{0\text{ (}1)}:\mathbf{1}$%
\end{tabular}
\\ \hline\hline
\end{tabular}%
\ \ 
\end{equation*}%
\caption{\textit{Fermionic content for three generations with\ }$\protect%
\beta =-\frac{1}{\protect\sqrt{3}}$\textit{\ (}$-\protect\sqrt{3}).$ \textit{%
The third component is written with its electric charge. We take }$m=1,2.$}
\label{tab:espectro}
\end{table}

For the scalar sector, we introduce the triplet field $\chi $ with Vacuum
Expectation Value (VEV) $\left\langle \chi \right\rangle ^{T}=\left( 0,0,\nu
_{\chi }\right) $, which induces the masses to the third fermionic
components. In the second transition it is necessary to introduce two
triplets$\;\rho $ and $\eta $ with VEV $\left\langle \rho \right\rangle
^{T}=\left( 0,\nu _{\rho },0\right) $ and $\left\langle \eta \right\rangle
^{T}=\left( \nu _{\eta },0,0\right) $ in order to give masses to the quarks
of type up and down respectively.

\begin{table}[tbp]
\begin{equation*}
\begin{tabular}{||c||c||c||}
\hline\hline
Representation $A$ & Representation $B$ & Representation $C$ \\ \hline\hline
$%
\begin{tabular}{c}
$q_{mL}=\left( 
\begin{array}{c}
d,s \\ 
-u,-c \\ 
J_{1},J_{2}%
\end{array}%
\right) _{L}:\mathbf{3}^{\ast }$ \\ 
$q_{3L}=\left( 
\begin{array}{c}
t \\ 
b \\ 
J_{3}%
\end{array}%
\right) _{L}:\mathbf{3}$%
\end{tabular}%
\ $ & $%
\begin{tabular}{c}
$q_{mL}=\left( 
\begin{array}{c}
d,b \\ 
-u,-t \\ 
J_{1},J_{3}%
\end{array}%
\right) _{L}:\mathbf{3}^{\ast }$ \\ 
$q_{3L}=\left( 
\begin{array}{c}
c \\ 
s \\ 
J_{2}%
\end{array}%
\right) _{L}:\mathbf{3}$%
\end{tabular}%
\ $ & $%
\begin{tabular}{c}
$q_{mL}=\left( 
\begin{array}{c}
s,b \\ 
-c,-t \\ 
J_{2},J_{3}%
\end{array}%
\right) _{L}:\mathbf{3}^{\ast }$ \\ 
$q_{3L}=\left( 
\begin{array}{c}
u \\ 
d \\ 
J_{1}%
\end{array}%
\right) _{L}:\mathbf{3}$%
\end{tabular}%
\ $ \\ \hline\hline
\end{tabular}%
\end{equation*}%
\caption{\textit{Three different assignments for the $SU(3)_L$ family
representation of quarks}}
\label{tab:combination}
\end{table}

In the gauge boson spectrum associated with the group $SU(3)_{L}\otimes
U(1)_{X},$ we are just interested in the physical neutral sector that
corresponds to the photon, $Z$ and $Z^{\prime },$ which are written in terms
of the electroweak basis for $\beta =-\frac{1}{\sqrt{3}}$ and $-\sqrt{3}$ as \cite{beta-arbitrario}

\begin{eqnarray}
A_{\mu } &=&S_{W}W_{\mu }^{3}+C_{W}\left( \beta T_{W}W_{\mu }^{8}+\sqrt{%
1-\beta ^{2}T_{W}^{2}}B_{\mu }\right) ,  \notag \\
Z_{\mu } &=&C_{W}W_{\mu }^{3}-S_{W}\left( \beta T_{W}W_{\mu }^{8}+\sqrt{%
1-\beta ^{2}T_{W}^{2}}B_{\mu }\right) ,  \notag \\
Z_{\mu }^{\prime } &=&-\sqrt{1-\beta ^{2}T_{W}^{2}}W_{\mu }^{8}+\beta
T_{W}B_{\mu },
\end{eqnarray}%
and with eigenvalues 
\begin{equation}
M_{A_{\mu }}^{2}=0;\quad M_{Z_{\mu }}^{2}\simeq \frac{g^{2}}{2}\left[ \frac{%
3g^{2}+4g^{\prime 2}}{3g^{2}+g^{\prime 2}}\right] \left( \nu _{\rho
}^{2}+\nu _{\eta }^{2}\right) ;\quad M_{Z_{\mu }^{\prime }}^{2}\simeq \frac{2%
\left[ 3g^{2}+g^{\prime 2}\right] }{9}\nu _{\chi }^{2},
\end{equation}%
where the Weinberg angle is defined as 
\begin{equation}
S_{W}=\sin \theta _{W}=\frac{g^{\prime }}{\sqrt{g^{2}+\left( 1+\beta
^{2}\right) g^{\prime 2}}},\quad T_{W}=\tan \theta _{W}=\frac{g^{\prime }}{%
\sqrt{g^{2}+\beta^{2}g^{\prime 2}}}
\end{equation}%
and $g,$ $g^{\prime }$ correspond to the coupling constants of the groups $%
SU(3)_{L}$ and $U(1)_{X}$ respectively. Further, a small mixing angle
between the two neutral currents $Z_{\mu }$ and $Z_{\mu }^{\prime }$ appears
with the following mass eigenstates

\begin{eqnarray}
Z_{1\mu } &=&Z_{\mu }C_{\theta }+Z_{\mu }^{\prime }S_{\theta };\quad Z_{2\mu
}=-Z_{\mu }S_{\theta }+Z_{\mu }^{\prime }C_{\theta };  \notag \\
\tan \theta &=&\frac{1}{\Lambda +\sqrt{\Lambda ^{2}+1}};\quad \Lambda =\frac{%
-2S_{W}C_{W}^{2}g^{\prime 2}\nu _{\chi }^{2}+\frac{3}{2}S_{W}T_{W}^{2}g^{2}%
\left( \nu _{\eta }^{2}+\nu _{\rho }^{2}\right) }{gg^{\prime }T_{W}^{2}\left[
3\beta S_{W}^{2}\left( \nu _{\eta }^{2}+\nu _{\rho }^{2}\right)
+C_{W}^{2}\left( \nu _{\eta }^{2}-\nu _{\rho }^{2}\right) \right] }.
\label{mix}
\end{eqnarray}

\section{Neutral currents\label{sec:neutral-currents}}

Using the fermionic content from table \ref{tab:espectro}, we obtain the
neutral coupling for the SM fermions

\begin{eqnarray}
\mathcal{L}^{NC} &=&\sum_{j=1}^{3}\left\{ \frac{g}{2C_{W}}\overline{\text{Q}%
_{j}}\gamma _{\mu }\left[ 2T_{3}P_{L}-2Q_{\text{Q}_{j}}S_{W}^{2}\right] 
\text{Q}_{j}Z^{\mu }\right.  \notag \\
&+&\frac{g}{2C_{W}}\overline{\ell _{j}}\gamma _{\mu }\left[
2T_{3}P_{L}-2Q_{\ell _{j}}S_{W}^{2}\right] \ell _{j}Z^{\mu }  \notag \\
&+&\left. \frac{g^{\prime }}{2T_{W}}\overline{\ell _{j}}\gamma _{\mu }\left[
\left( -2T_{8}-\beta T_{W}^{2}\Lambda _{3}\right) P_{L}+2\beta Q_{\ell
_{j}}T_{W}^{2}P_{R}\right] \ell _{j}Z^{\mu \prime }\right\}  \notag \\
&+&\sum_{m=1}^{2}\frac{g^{\prime }}{2T_{W}}\overline{q_{m}}\gamma _{\mu }%
\left[ \left( 2T_{8}+\beta Q_{q_{m}}T_{W}^{2}\Lambda _{1}\right)
P_{L}+2\beta Q_{q_{m}}T_{W}^{2}P_{R}\right] q_{m}Z^{\mu \prime }  \notag \\
&+&\frac{g^{\prime }}{2T_{W}}\overline{q_{3}}\gamma _{\mu }\left[ \left(
-2T_{8}+\beta Q_{q_{3}}T_{W}^{2}\Lambda _{2}\right) P_{L}+2\beta
Q_{q_{3}}T_{W}^{2}P_{R}\right] q_{3}Z^{\mu \prime },  \label{lag-1}
\end{eqnarray}%
where Q$_{j}$ with $j=1,2,3$ has been written in a SM-like notation i.e. it
refers to triplets of quarks associated with the three generations of quarks
(SM does not make difference in the family representations). On the other
hand, the coupling of the exotic gauge boson ($Z_{\mu }^{\prime }$) with the
two former families are different from the ones involving the third family.
This is because the third familiy transforms differently as it was remarked
in table \ref{tab:espectro}. Consequently, there are terms where only the
components $m=1,2$ are summed, leaving the third one in a term apart. $%
Q_{q_{j}}$ are the electric charges. The Gell-Mann matrices $T_{3}$ $=\frac{1%
}{2}diag(1,-1,0)$ and $T_{8}=\frac{1}{2\sqrt{3}}diag(1,1,-2)$ are introduced
in the notation. We also define $\Lambda _{1}=diag(-1,\frac{1}{2},2)$, $%
\Lambda _{2}=diag(\frac{1}{2},-1,2)$ and the projectors $P_{R,L}=\frac{1}{2}%
(1\pm \gamma _{5}).$ Finally, $\ell _{j}$ denote the leptonic triplets with $%
Q_{\ell _{j}}$ denoting their electric charges and $\Lambda
_{3}=diag(1,1,2Q_{1})$ with $Q_{1}=0,-1$ for $\beta =-\frac{1}{\sqrt{3}},-%
\sqrt{3}$ respectively$.$

The neutral lagrangian (\ref{lag-1}) can be written as

\begin{eqnarray}
\mathcal{L}^{NC} &=&\frac{g}{2C_{W}}\left\{ \sum_{j=1}^{3}\overline{\text{Q}%
_{j}}\gamma _{\mu }\left[ g_{V}^{Q_{j}}-g_{A}^{Q_{j}}\gamma _{5}\right] 
\text{Q}_{j}Z^{\mu }+\overline{\ell _{j}}\gamma _{\mu }\left[ g_{V}^{\ell
_{j}}-g_{A}^{\ell _{j}}\gamma _{5}\right] \ell _{j}Z^{\mu }\right.  \notag \\
&&+\overline{\ell _{j}}\gamma _{\mu }\left[ \overset{\sim }{g}_{V}^{\ell
_{j}}-\overset{\sim }{g}_{A}^{\ell _{j}}\gamma _{5}\right] \ell _{j}Z^{\mu
\prime }+\sum_{m=1}^{2}\overline{q_{m}}\gamma _{\mu }\left[ \overset{\sim }{g%
}_{V}^{q_{m}}-\overset{\sim }{g}_{A}^{q_{m}}\gamma _{5}\right] q_{m}Z^{\mu
\prime }  \notag \\
&&\left. +\overline{q_{3}}\gamma _{\mu }\left[ \overset{\sim }{g}%
_{V}^{q_{3}}-\overset{\sim }{g}_{A}^{q_{3}}\gamma _{5}\right] q_{3}Z^{\mu
\prime }\right\} ,  \label{lag-2}
\end{eqnarray}
with the vector and axial vector couplings given by

\begin{eqnarray}
g_{V}^{f} &=&T_{3}-2Q_{f}S_{W}^{2},\qquad g_{A}^{f}=T_{3}  \notag \\
\overset{\sim }{g}_{V,A}^{q_{m}} &=&\frac{C_{W}^{2}}{\sqrt{1-\left( 1+\beta
^{2}\right) S_{W}^{2}}}\left[ T_{8}+\beta Q_{q_{m}}T_{W}^{2}\left( \frac{1}{2%
}\Lambda _{1}\pm 1\right) \right]   \notag \\
\overset{\sim }{g}_{V,A}^{q_{3}} &=&\frac{C_{W}^{2}}{\sqrt{1-\left( 1+\beta
^{2}\right) S_{W}^{2}}}\left[ -T_{8}+\beta Q_{q_{3}}T_{W}^{2}\left( \frac{1}{%
2}\Lambda _{2}\pm 1\right) \right]   \notag \\
\overset{\sim }{g}_{V,A}^{\ell _{j}} &=&\frac{C_{W}^{2}}{\sqrt{1-\left(
1+\beta ^{2}\right) S_{W}^{2}}}\left[ -T_{8}-\beta T_{W}^{2}\left( \frac{1}{2%
}\Lambda _{3}\mp Q_{\ell _{j}}\right) \right] ,  \label{coup1}
\end{eqnarray}%
where $f=$Q$_{j},$ $\ell _{j}$ in the first line. It is noted that $%
g_{V,A}^{f}$ are the same as the SM definitions and $\overset{\sim }{g}%
_{V,A}^{f}$are new $\beta $-dependent couplings of $Z_{\mu }^{\prime }$
(i.e. model dependent), whose values also depend on the family realization
from table \ref{tab:combination}. On the other hand, there is a small mixing
angle between $Z_{\mu }$ and $Z_{\mu }^{\prime }$ given by Eq. (\ref{mix}),
where $Z_{1\mu }$ is the SM-like neutral boson and $Z_{2\mu }$ the exotic
ones. Taking a very small angle, we can do C$_{\theta }\simeq 1$ so that the
lagrangian (\ref{lag-2}) becomes 
\begin{eqnarray}
\mathcal{L}^{NC} &=&\sum_{j=1}^{3}\left\{ \frac{g}{2C_{W}}\overline{%
\text{Q}_{j}}\gamma _{\mu }\left[ G_{V}^{Q_{j}}-G_{A}^{Q_{j}}\gamma _{5}%
\right] \text{Q}_{j}Z_{1}^{\mu }+\frac{g}{2C_{W}}\overline{\ell _{j}}\gamma
_{\mu }\left[ G_{V}^{\ell _{j}}-G_{A}^{\ell _{j}}\gamma _{5}\right] \ell
_{j}Z_{1}^{\mu }\right.   \notag \\
&&+\left. \frac{g}{2C_{W}}\overline{\text{Q}_{j}}\gamma _{\mu }\left[ 
\overset{\sim }{G}_{V}^{Q_{j}}-\overset{\sim }{G}_{A}^{Q_{j}}\gamma _{5}%
\right] \text{Q}_{j}Z_{2}^{\mu }+\frac{g}{2C_{W}}\overline{\ell _{j}}\gamma
_{\mu }\left[ \overset{\sim }{G}_{V}^{\ell _{j}}-\overset{\sim }{G}%
_{A}^{\ell _{j}}\gamma _{5}\right] \ell _{j}Z_{2}^{\mu }\right\} ,
\label{lag-3}
\end{eqnarray}%
where the couplings associated with $Z_{1\mu }$ are

\begin{eqnarray}
G_{V,A}^{f} &=&g_{V,A}^{f}+\delta g_{V,A}^{f},  \notag  \label{coupling-2} \\
\delta g_{V,A}^{f} &=&\overset{\sim }{g}_{V,A}^{f}S_{\theta },  \label{coup2}
\end{eqnarray}%
and the couplings associated with $Z_{2\mu }$ are

\begin{eqnarray}
\overset{\sim }{G}_{V,A}^{f} &=&\overset{\sim }{g}_{V,A}^{f}-\delta \overset{%
\sim }{g}_{V,A}^{f},  \notag \\
\delta \overset{\sim }{g}_{V,A}^{f} &=&g_{V,A}^{f}S_{\theta }.  \label{coup3}
\end{eqnarray}

\section{Z-Pole Observables\label{sec:z-pole}}

The couplings of the $Z_{1\mu }$ in eq. (\ref{lag-3}) have the same form as
the SM neutral couplings but by replacing the vector and axial vector
couplings $g_{V,A}^{SM}$ by $G_{V,A}=g_{V,A}^{SM}+\delta g_{V,A},$ where $%
\delta g_{V,A}$ (given by eq. (\ref{coup2})) is a correction due to the
small $Z_{\mu }-Z_{\mu }^{\prime }$ mixing angle $\theta .$ For this reason
all the analytical parameters at the Z pole have the same SM-form but with
small correction factors that depend on the family assignment. The partial
decay widths of $Z_{1}$ into fermions $f\overline{f}$ is described by \cite%
{one, pitch}:

\begin{equation}
\Gamma _{f}^{SM}=\frac{N_{c}^{f}G_{f}M_{Z_{1}}^{3}}{6\sqrt{2}\pi }\rho _{f}%
\left[ \frac{3\beta _{K}-\beta _{K}^{3}}{2}\left( g_{V}^{f}\right)
^{2}+\beta _{K}^{3}\left( g_{A}^{f}\right) ^{2}\right] R_{QED}R_{QCD},
\label{partial-decay}
\end{equation}

\noindent where $N_{c}^{f}=1$, 3 for leptons and quarks respectively, $%
R_{QED,QCD}$ are global final-state QED and QCD corrections, and $\beta _{K}=%
\sqrt{1-\frac{4m_{b}^{2}}{M_{Z}^{2}}}$ considers kinematic corrections only
important for the $b$-quark. Universal electroweak corrections sensitive to
the top quark mass are taken into account in $\rho _{f}=1+\rho _{t}$ and in $%
g_{V}^{SM}$ which is written in terms of an effective Weinberg angle \cite%
{one}

\begin{equation}
\overline{S_{W}}^{2}=\kappa _{f}S_{W}^{2}=\left( 1+\frac{\rho _{t}}{T_{W}^{2}%
}\right) S_{W}^{2},  \label{effective-angle}
\end{equation}

\noindent with $\rho _{t}=3G_{f}m_{t}^{2}/8\sqrt{2}\pi ^{2}$. Non-universal
vertex corrections are also taken into account in the $Z_{1}\overline{b}b$
vertex with additional one-loop leading terms given by \cite{one, pitch}

\begin{equation}
\rho _{b}\rightarrow \rho _{b}-\frac{4}{3}\rho _{t}\text{ and }\kappa
_{b}\rightarrow \kappa _{b}+\frac{2}{3}\rho _{t}.  \label{Zb-vertex}
\end{equation}

Table \ref{tab:observables} resumes some observables, with their
experimental values from CERN collider (LEP), SLAC Liner Collider (SLC) and
data from atomic parity violation \cite{one}, the SM predictions and the
expressions predicted by 331 models. We use $M_{Z_{1}}=91.1876$ $GeV,$ $%
m_{t}=176.9$ $GeV$, $S_{W}^{2}=0.2314$, and for $m_{b}$ we use \cite{greub}

\begin{equation*}
\overline{m}_{b}(\mu \rightarrow M_{Z_{1}})=m_{b}\left[ 1+\frac{\alpha
_{S}(\mu )}{\pi }\left( \ln \frac{m_{b}^{2}}{\mu ^{2}}-\frac{4}{3}\right) %
\right] ,
\end{equation*}

\noindent with $m_{b}\approx 4.5$ $GeV$ the pole mass, $\overline{m}_{b}(\mu
\rightarrow M_{Z_{1}})$ the running mass at $M_{Z_{1}}$ scale in the $%
\overline{MS}$ scheme, and $\alpha _{S}(M_{Z_{1}})=0.1213\pm 0.0018$ the
strong coupling constant.

The 331 predictions from table \ref{tab:observables} are expressed in terms
of SM values corrected by

\begin{eqnarray}
\delta _{Z} &=&\frac{\Gamma _{u}^{SM}}{\Gamma _{Z}^{SM}}(\delta _{u}+\delta
_{c})+\frac{\Gamma _{d}^{SM}}{\Gamma _{Z}^{SM}}(\delta _{d}+\delta _{s})+%
\frac{\Gamma _{b}^{SM}}{\Gamma _{Z}^{SM}}\delta _{b}+3\frac{\Gamma _{\nu
}^{SM}}{\Gamma _{Z}^{SM}}\delta _{\nu }+3\frac{\Gamma _{e}^{SM}}{\Gamma
_{Z}^{SM}}\delta _{\ell };  \notag \\
\delta _{had} &=&R_{c}^{SM}(\delta _{u}+\delta _{c})+R_{b}^{SM}\delta _{b}+%
\frac{\Gamma _{d}^{SM}}{\Gamma _{had}^{SM}}(\delta _{d}+\delta _{s});  \notag
\\
\delta _{\sigma } &=&\delta _{had}+\delta _{\ell }-2\delta _{Z};  \notag \\
\delta A_{f} &=&\frac{\delta g_{V}^{f}}{g_{V}^{f}}+\frac{\delta g_{A}^{f}}{%
g_{A}^{f}}-\delta _{f},  \label{shift1}
\end{eqnarray}

\noindent where for the light fermions

\begin{equation}
\delta _{f}=\frac{2g_{V}^{f}\delta g_{V}^{f}+2g_{A}^{f}\delta g_{A}^{f}}{%
\left( g_{V}^{f}\right) ^{2}+\left( g_{A}^{f}\right) ^{2}},  \label{shift2}
\end{equation}

\noindent while for the $b$-quark

\begin{equation}
\delta _{b}=\frac{\left( 3-\beta _{K}^{2}\right) g_{V}^{b}\delta
g_{V}^{b}+2\beta _{K}^{2}g_{A}^{b}\delta g_{A}^{b}}{\left( \frac{3-\beta
_{K}^{2}}{2}\right) \left( g_{V}^{b}\right) ^{2}+\beta _{K}^{2}\left(
g_{A}^{b}\right) ^{2}}.  \label{shift3}
\end{equation}

\noindent The above expressions are evaluated in terms of the effective
Weinberg angle from eq. (\ref{effective-angle}).

\begin{table}[tbp]
\begin{center}
$%
\begin{tabular}{|c|c|c|c|}
\hline
Quantity & Experimental Values & Standard Model & 331 Model \\ \hline
$\Gamma _{Z}$ $\left[ GeV\right] $ & 2.4952 $\pm $ 0.0023 & 2.4972 $\pm $
0.0012 & $\Gamma _{Z}^{SM}\left( 1+\delta _{Z}\right) $ \\ \hline
$\Gamma _{had}$ $\left[ GeV\right] $ & 1.7444 $\pm $ 0.0020 & 1.7435 $\pm $
0.0011 & $\Gamma _{had}^{SM}\left( 1+\delta _{had}\right) $ \\ \hline
$\Gamma _{\left( \ell ^{+}\ell ^{-}\right) }$ $MeV$ & 83.984 $\pm $ 0.086 & 
84.024 $\pm $ 0.025 & $\Gamma _{\left( \ell ^{+}\ell ^{-}\right)
}^{SM}\left( 1+\delta _{\ell }\right) $ \\ \hline
$\sigma _{had}$ $\left[ nb\right] $ & 41.541 $\pm $ 0.037 & 41.472 $\pm $
0.009 & $\sigma _{had}^{SM}\left( 1+\delta _{\sigma }\right) $ \\ \hline
$R_{e}$ & 20.804 $\pm $ 0.050 & 20.750 $\pm $ 0.012 & $R_{e}^{SM}\left(
1+\delta _{had}+\delta _{e}\right) $ \\ \hline
$R_{\mu }$ & 20.785 $\pm $ 0.033 & 20.751 $\pm $ 0.012 & $R_{\mu
}^{SM}\left( 1+\delta _{had}+\delta _{\mu }\right) $ \\ \hline
$R_{\tau }$ & 20.764 $\pm $ 0.045 & 20.790 $\pm $ 0.018 & $R_{\tau
}^{SM}\left( 1+\delta _{had}+\delta _{\tau }\right) $ \\ \hline
$R_{b}$ & 0.21638 $\pm $ 0.00066 & 0.21564 $\pm $ 0.00014 & $%
R_{b}^{SM}\left( 1+\delta _{b}-\delta _{had}\right) $ \\ \hline
$R_{c}$ & 0.1720 $\pm $ 0.0030 & 0.17233 $\pm $ 0.00005 & $R_{c}^{SM}\left(
1+\delta _{c}-\delta _{had}\right) $ \\ \hline
$A_{e}$ & 0.15138 $\pm $ 0.00216 & 0.1472 $\pm $ 0.0011 & $A_{e}^{SM}\left(
1+\delta A_{e}\right) $ \\ \hline
$A_{\mu }$ & 0.142 $\pm $ 0.015 & 0.1472 $\pm $ 0.0011 & $A_{\mu
}^{SM}\left( 1+\delta A_{\mu }\right) $ \\ \hline
$A_{\tau }$ & 0.136 $\pm $ 0.015 & 0.1472 $\pm $ 0.0011 & $A_{\tau
}^{SM}\left( 1+\delta A_{\tau }\right) $ \\ \hline
$A_{b}$ & 0.925 $\pm $ 0.020 & 0.9347 $\pm $ 0.0001 & $A_{b}^{SM}\left(
1+\delta A_{b}\right) $ \\ \hline
$A_{c}$ & 0.670 $\pm $ 0.026 & 0.6678 $\pm $ 0.0005 & $A_{c}^{SM}\left(
1+\delta A_{c}\right) $ \\ \hline
$A_{s}$ & 0.895 $\pm $ 0.091 & 0.9357 $\pm $ 0.0001 & $A_{s}^{SM}\left(
1+\delta A_{s}\right) $ \\ \hline
$A_{FB}^{\left( 0,e\right) }$ & 0.0145 $\pm $ 0.0025 & 0.01626 $\pm $ 0.00025
& $A_{FB}^{(0,e)SM}\left( 1+2\delta A_{e}\right) $ \\ \hline
$A_{FB}^{\left( 0,\mu \right) }$ & 0.0169 $\pm $ 0.0013 & 0.01626 $\pm $
0.00025 & $A_{FB}^{(0,\mu )SM}\left( 1+\delta A_{e}+\delta A_{\mu }\right) $
\\ \hline
$A_{FB}^{\left( 0,\tau \right) }$ & 0.0188 $\pm $ 0.0017 & 0.01626 $\pm $
0.00025 & $A_{FB}^{(0,\tau )SM}\left( 1+\delta A_{e}+\delta A_{\tau }\right) 
$ \\ \hline
$A_{FB}^{\left( 0,b\right) }$ & 0.0997 $\pm $ 0.0016 & 0.1032 $\pm $ 0.0008
& $A_{FB}^{(0,b)SM}\left( 1+\delta A_{e}+\delta A_{b}\right) $ \\ \hline
$A_{FB}^{\left( 0,c\right) }$ & 0.0706 $\pm $ 0.0035 & 0.0738 $\pm $ 0.0006
& $A_{FB}^{(0,c)SM}\left( 1+\delta A_{e}+\delta A_{c}\right) $ \\ \hline
$A_{FB}^{\left( 0,s\right) }$ & 0.0976 $\pm $ 0.0114 & 0.1033 $\pm $ 0.0008
& $A_{FB}^{(0,s)SM}\left( 1+\delta A_{e}+\delta A_{s}\right) $ \\ \hline
$Q_{W}(Cs)$ & $-$72.69 $\pm $ 0.48 & $-$73.19 $\pm $ 0.03 & $%
Q_{W}^{SM}\left( 1+\delta Q_{W}\right) $ \\ \hline
\end{tabular}%
\ \ $%
\end{center}
\caption{\textit{The parameters for experimental values, SM predictions and
331 corrections. The values are taken from ref. \protect\cite{one}}}
\label{tab:observables}
\end{table}

For the predicted SM partial decay given by (\ref{partial-decay}), we use
the following values taken from ref. \cite{one}

\begin{eqnarray*}
\Gamma _{u}^{SM} &=&0.3004\pm 0.0002\text{ }GeV;\quad \Gamma
_{d}^{SM}=0.3832\pm 0.0002\text{ }GeV; \\
\Gamma _{b}^{SM} &=&0.3758\pm 0.0001\text{ }GeV;\quad \Gamma _{\nu
}^{SM}=0.16729\pm 0.00007\text{ }GeV; \\
\Gamma _{e}^{SM} &=&0.08403\pm 0.00004\text{ }GeV.
\end{eqnarray*}

The weak charge is written as

\begin{equation}
Q_{W}=Q_{W}^{SM}+\Delta Q_{W}=Q_{W}^{SM}\left( 1+\delta Q_{W}\right) ,
\label{weak}
\end{equation}
where $\delta Q_{W}=\frac{\Delta Q_{W}}{Q_{W}^{SM}}$. The deviation $\Delta
Q_{W}$ is \cite{cesio} 
\begin{equation}
\Delta Q_{W}=\left[ \left( 1+4\frac{S_{W}^{4}}{1-2S_{W}^{2}}\right) Z-N%
\right] \Delta \rho _{M}+\Delta Q_{W}^{\prime },  \label{dev}
\end{equation}
and $\Delta Q_{W}^{\prime }$ which contains new physics gives

\begin{eqnarray}
\Delta Q_{W}^{\prime } &=&-16\left[ \left( 2Z+N\right) \left( g_{A}^{e}%
\overset{\sim }{g}_{V}^{u}+\overset{\sim }{g}_{A}^{e}g_{V}^{u}\right)
+\left( Z+2N\right) \left( g_{A}^{e}\overset{\sim }{g}_{V}^{d}+\overset{\sim 
}{g}_{A}^{e}g_{V}^{d}\right) \right] S_{\theta }  \notag \\
&&-16\left[ \left( 2Z+N\right) \overset{\sim }{g}_{A}^{e}\overset{\sim }{g}%
_{V}^{u}+\left( Z+2N\right) \overset{\sim }{g}_{A}^{e}\overset{\sim }{g}%
_{V}^{d}\right] \frac{M_{Z_{1}}^{2}}{M_{Z_{2}}^{2}}.  \label{new}
\end{eqnarray}

For Cesium we have $Z=55,$ $N=78$, and for the first term in (\ref{dev}) we
take the value $\left[ \left( 1+4\frac{S_{W}^{4}}{1-2S_{W}^{2}}\right) Z-N%
\right] \Delta \rho _{M}\simeq -0.01$ \cite{cesio}. With the definitions in
eq. (\ref{coup1}) for $\beta =-\frac{1}{\sqrt{3}}$ and $-\sqrt{3}$, we
displays in table \ref{tab:weak-charge} the new physics corrections to $%
\Delta Q_{W}$ given by eq. (\ref{new}) for each representation of quarks
listed in table \ref{tab:combination}. We get the same correction for the
spectrum $A$ and $B$ due to the fact that the weak charge depends mostly on
the up-down quarks, and $A,B$-cases maintain the same representation for
this family.

\begin{table}[tbp]
\begin{center}
$ 
\begin{tabular}{|c|c|}
\hline
& $\beta =-\frac 1{\sqrt{3}}$ \\ \hline
$A,B$ & $\Delta Q_W^{\prime }=\left( 4.68Z+2.98N\right) S_\theta +\left(
3.11Z+4.00N\right) \frac{M_{Z_1}^2}{M_{Z_2}^2}$ \\ \hline
$C$ & $\Delta Q_W^{\prime }=-\left( 8.125Z+9.82N\right) S_\theta -\left(
5.78Z+4.89N\right) \frac{M_{Z_1}^2}{M_{Z_2}^2}$ \\ \hline
& $\beta =-\sqrt{3}$ \\ \hline
$A,B$ & $\Delta Q_W^{\prime }=\left( 2.81Z+1.26N\right) S_\theta +\left(
5.85Z+42.25N\right) \frac{M_{Z_1}^2}{M_{Z_2}^2}$ \\ \hline
$C$ & $\Delta Q_W^{\prime }=-\left( 36.23Z+37.785N\right) S_\theta -\left(
115.04Z+78.645N\right) \frac{M_{Z_1}^2}{M_{Z_2}^2}$ \\ \hline
\end{tabular}
$%
\end{center}
\caption{\textit{New physics contributions to }$\Delta Q_{W}$ \textit{for
the two 331 models according to the family assignment from table }  \protect
\ref{tab:combination}.}
\label{tab:weak-charge}
\end{table}
With the expressions for the Z-pole observables and the experimental data
shown in table \ref{tab:observables}, we perform a $\chi ^{2}$ fit for each
representation $A,B$ and $C$ at 95\% CL, which will allow us to display the
family dependence in the model. We find the best allowed region in the plane 
$M_{Z_{2}}-S_{\theta }.$ Figure 1 displays three cases for $\beta =-\frac{1}{%
\sqrt{3}}$, each one corresponding to the family representations from table %
\ref{tab:combination} respectively, exhibiting family-dependent regions. The
bounds for $M_{Z_{2}}$ and $S_{\theta }$ are shown in table \ref{tab:bound-1}
for each family-choices. We can see that the lowest bound for $M_{Z_{2}}$ is
about $1400$ $GeV$ for models $A$ and $B$, while model $C$ with the lightest
family defined in the third multiplet increases this bound to 2100 $GeV.$ It
is also noted that the $A$ and $B$ representations yield broader allowed
regions for the mixing angle that model $C,$ showing that the family choices
is a fundamental issue in the phenomenology of 331 models. $A$-region and $B$%
-region are very similar because they present the same weak corrections, as
it is shown in table \ref{tab:weak-charge}; the small differences arise
mostly due to the bottom correction in eq. (\ref{shift3}). The table \ref%
{tab:weak-charge} also shows that the $C$-region contribute to the new
physics correction with a contrary sign respect $A$ and $B$ corrections.

\begin{table}[tbp]
\begin{center}
\begin{tabular}{|c|c|c|}
\hline\hline
Quarks Rep. & $M_{Z_{2}}$ ($GeV$) & $S_{\theta }$ ($\times 10^{-3}$) \\ 
\hline
Rep. $A$ & $M_{Z_{2}}\geq 1400$ & $-0.9\leq S_{\theta }\leq 2$ \\ \hline
Rep. $B$ & $M_{Z_{2}}\geq 1400$ & $-0.9\leq S_{\theta }\leq 2$ \\ \hline
Rep. $C$ & $M_{Z_{2}}\geq 2100$ & $-0.9\leq S_{\theta }\leq 0.9$ \\ \hline
\end{tabular}%
\end{center}
\caption{\textit{Bounds with }$\protect\beta =-\frac{1}{\protect\sqrt{3}}$%
\textit{\ for M}$_{Z_{2}}$ \textit{and S}$_{\protect\theta }$ \textit{for
three quark representations} \textit{at 95\% CL}}
\label{tab:bound-1}
\end{table}
For $\beta =-\sqrt{3}$, we get the regions in figure 2, which also compares
the three family choices. This model increases the lowest bound in the $%
M_{Z_{2}}$ value to 4000 $GeV$ for the $A$ and $B$ spectrums, and to 10000 $%
GeV$ for the $C$ spectrum$.$ In this case $A$ and $B$ regions are
equivalent. The table \ref{tab:bound-2} displays the bounds for the mixing
angle and $M_{Z_{2}}$. We can see that the mixing angle is smaller in about
one order of magnitude that the angle predicted by $\beta =-\frac{1}{\sqrt{3}%
}$ model$.$ These differences between both models arise mostly because they
present variations to the weak charge corrections (see table \ref%
{tab:weak-charge}).

\begin{table}[tbp]
\begin{center}
\begin{tabular}{|c|c|c|}
\hline\hline
Quarks Rep. & $M_{Z_{2}}$ ($GeV$) & $S_{\theta }$ ($\times 10^{-4}$) \\ 
\hline
Rep. $A$ & $M_{Z_{2}}\geq 4000$ & $-1.2\leq S_{\theta }\leq 1.7$ \\ \hline
Rep. $B$ & $M_{Z_{2}}\geq 4000$ & $-1.2\leq S_{\theta }\leq 1.7$ \\ \hline
Rep. $C$ & $M_{Z_{2}}\geq 10000$ & $-1.2\leq S_{\theta }\leq 1.2$ \\ \hline
\end{tabular}%
\end{center}
\caption{\textit{Bounds with }$\protect\beta =-\protect\sqrt{3}$\textit{\
for M}$_{Z_{2}}$ \textit{and S}$_{\protect\theta }$ \textit{for three quark
representations} \textit{at 95\% CL}}
\label{tab:bound-2}
\end{table}

\section{Conclusions\label{conclusions}}

The $SU(3)_{c}\otimes $ $SU(3)_{L}\otimes U(1)_{X}$ models for three
families with $\beta =-\frac{1}{\sqrt{3}}$ and $-\sqrt{3},$ corresponding to
the Long \cite{twelve} and Pisano-Pleitez-Frampton bilepton models \cite{ten}
respectively, were studied under the framework of family dependence.

As it is shown in table \ref{tab:combination}, we found three different
assignments of quarks into the mass family basis. Each assignment determines
different weak couplings of the quarks to the extra neutral current
associated to $Z_{2}$, which holds a small angle mixing with respect to the
SM-neutral current associated to $Z_{1}.$ This mixing gives different
allowed regions in the $M_{Z_{2}}-S_{\theta }$ plane for the LEP parameters
at the Z-pole and including data from the atomic parity violation.

Performing a $\chi ^{2}$ fit at 95\% CL we found for the Long model that $%
M_{Z_{2}}\geq 1400$ when $-0.9\times 10^{-3}\leq S_{\theta }\leq 2\times
10^{-3}$ for $A$ and $B$ assignments, and $M_{Z_{2}}\geq 2100$ when $%
-0.9\times 10^{-3}\leq S_{\theta }\leq 0.9\times 10^{-3}$ for $C$
assignment. For the Pleitez model we got $M_{Z_{2}}\geq 4000$ for $%
-1.2\times 10^{-4}\leq S_{\theta }\leq 1.7\times 10^{-4}$, and $%
M_{Z_{2}}\geq 10000$ for $-1.2\times 10^{-4}\leq S_{\theta }\leq 1.2\times
10^{-4}$ for each case.

Unlike the SM where the family assignment is arbitrary without any
phenomenological change, our results show how this assignment yields
differences in the numerical predictions for two 331 models. We see that the
lowest bound for $M_{Z_{2}}$ is higher than those obtained by other authors
for one family models \cite{fourteen}. Due to the restriction of the data
from the atomic parity violation, we are getting a differences of about one
order of magnitude in the lowest bound for the $M_{Z_{2}}$. In addition we
found that the bounds associated with the angle mixing is highly suppressed (%
$\sim $ $10^{-4}$) in the Pleitez model when the lightest quarks family
transform differently respect the two heavier families.

This study can be extended if we consider linear combinations among the
three familiy assignments according to the ansatz of the quarks mass matrix
in agreement with the physical mass and mixing angle mass. In this case, the
allowed regions for $M_{Z_{2}}$ Vs $S_{\theta }$ would be a combination
among the regions obtained for $A$, $B$ and $C$ models.

We acknowledge the financial support from COLCIENCIAS.

\newpage

\begin{figure}[tbph]
\centering \includegraphics[scale=0.8]{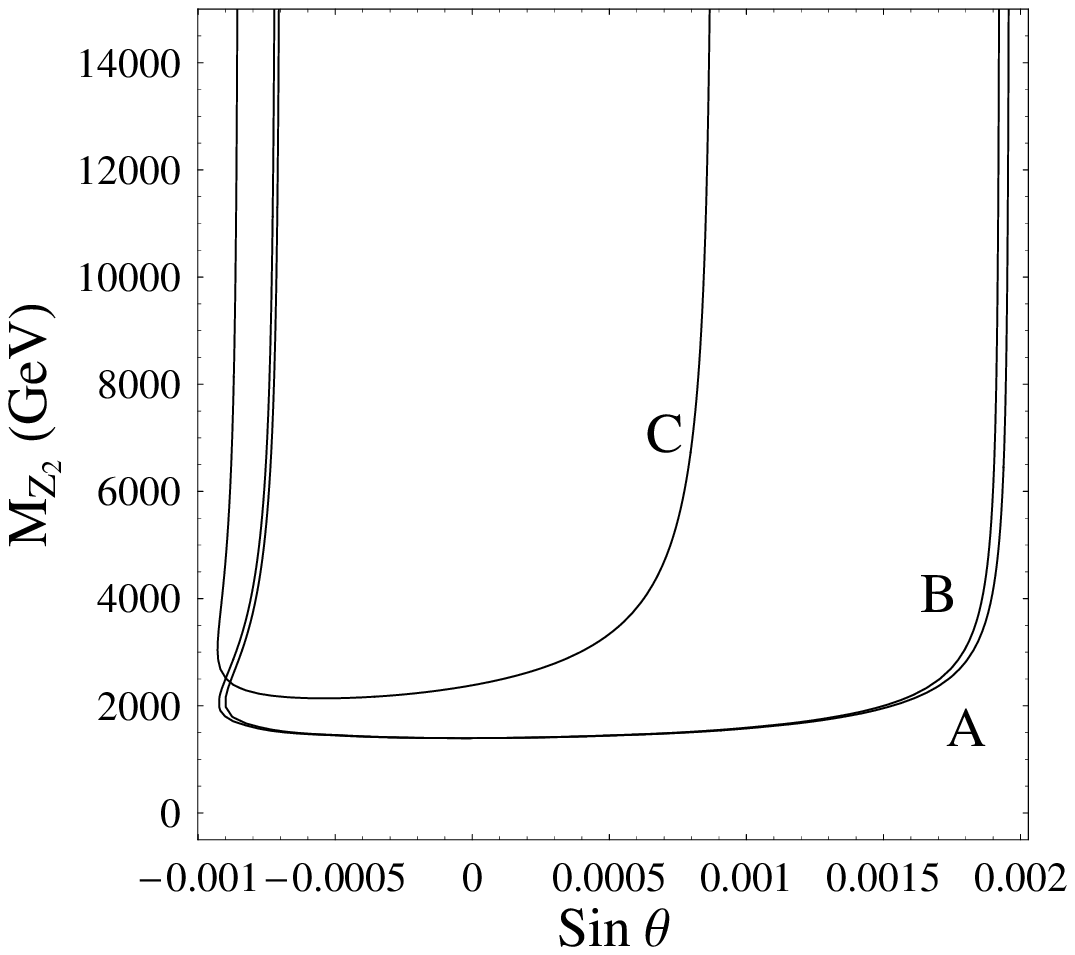}
\caption{\textit{The allowed region for }$M_{Z_{2}}\;vs\;\sin \protect\theta 
$\textit{\ in the model with }$\protect\beta =-1/\protect\sqrt{3}$\textit{.
A, B and C correspond to the assignment of families from table 2}}
\label{figura1}
\end{figure}

\newpage

\begin{figure}[tbph]
\centering \includegraphics[scale=0.8]{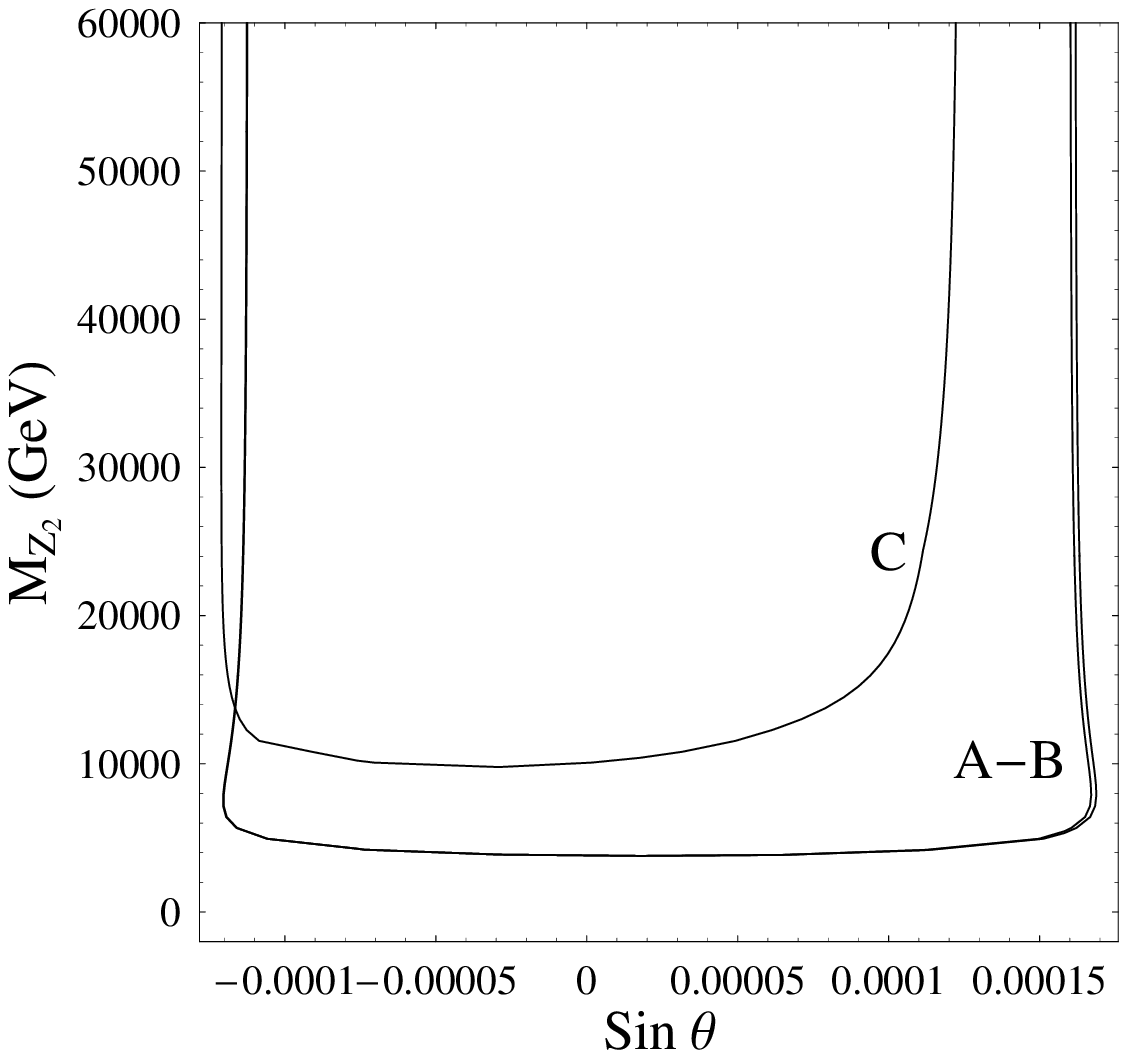}
\caption{\textit{The allowed region for }$M_{Z_{2}}\;vs\;\sin \protect\theta 
$\textit{\ in the model with }$\protect\beta =-\protect\sqrt{3}$\textit{. A,
B and C correspond to the assignment of families from table 2}}
\label{figura2}
\end{figure}

\end{document}